\documentstyle[preprint,aps]{revtex}
\begin{document}
\draft
\title{CROSSOVER BETWEEN THE DENSE ELECTRON-HOLE 
 PHASE AND THE BCS EXCITONIC PHASE IN QUANTUM DOTS}
\author{BORIS A. RODRIGUEZ$^1$\cite{boris},
 AUGUSTO GONZALEZ$^{2}$\cite{augusto}, 
 LUIS QUIROGA$^3$\cite{luis},  
 ROBERTO CAPOTE$^4$\cite{capote}, and 
 FERNEY J. RODRIGUEZ$^3$\cite{ferney}}
\address{ $^1$Departamento de Fisica, Universidad de Antioquia,
 AA 1226, Medellin, Colombia\\
 $^2$Instituto de Cibernetica, Matematica y Fisica Calle E 
 309, Vedado, Habana 4, Cuba\\ 
 $^3$Departamento de Fisica, Universidad de los Andes, AA 4976, 
 Bogota, Colombia\\
 $^4$Centro de Estudios Aplicados al Desarrollo Nuclear, Calle
 30 No 502, Miramar, La Habana, Cuba}
\date{\today}

\maketitle

\begin{abstract}
Second order perturbation theory and a Lipkin-Nogami scheme combined 
with an exact Monte Carlo projection after 
variation are applied to compute the ground-state energy of 
$6\le N\le 210$ electron-hole pairs confined in a parabolic 
two-dimensional quantum dot. The energy shows nice scaling properties as 
$N$ or the confinement strength is varied. A crossover from the 
high-density electron-hole phase to the BCS excitonic phase is found
at a density which is roughly four times the close-packing density
of excitons. 
\end{abstract}

\newpage

\section{Introduction}

As we understand, the interest in electron-hole states (excitons) 
in semiconductor physics is motivated by two facts. First, 
excitons have a bosonic character (as they are made up of a 
pair of fermions) and, thus, the many-exciton system is a
candidate for a Bose condensate. This possibility was envisaged
long ago \cite{Keldysh}, but regained attention in the last
years after the Bose condensation of alcali atoms was 
achieved \cite{bec}. Experimentally, signals of Bose-Einstein
statistics have been identified in the photoluminiscence of 
quantum wells under strong laser pumping \cite{KW98}, and 
indirect excitons in quantum wells are being manipulated via
applied stress and inhomogeneous electric fields \cite{NSE99} 
to reach the densities needed for Bose condensation. On the
other hand, excitons are at the basis of many optical properties
of semiconductors \cite{Bastard}. Recent experimental works have
focused on the lowest dimensional structures, and very 
interesting properties have been found in the photoluminiscence 
of quantum wires \cite{qwires} and single quantum dots 
\cite{experiments}. 

In the present paper, we study a two-dimensional quantum 
dot with a number of electron-hole pairs, $6\le N\le 210$,
i. e. intermediate between the very small dot
\cite{experiments} and the quantum well \cite{KW98}. We 
study the dot at strong and intermediate confinement regimes
by means of second-order perturbation theory and a variational
(BCS) procedure. The main results of the paper may be summarised 
as follows.

We found a breakdown of perturbation theory and a significant
BCS pairing roughly at the same confinement strength,
corresponding approximately to four times the close-packing
density of excitons (or four times the Mott transition density).
We notice that the BCS calculations were performed within the
Lipkin-Nogami scheme \cite{LN} with exact projection onto the
$N$-pair sector \cite{CG} to avoid the incorrect behaviour of 
the naive BCS function in a finite system \cite{BCSspu}.  
Second, we found that the energy depends on $N$ and the 
confinement strength in a scaled way.

Our paper is complementary to \onlinecite{GC98,PG99}, in which the
multiexcitonic quantum dot is also studied. In \onlinecite{GC98}, 
we found that the far-infrared absorption of the dot is 
dominated by a giant-dipole resonance similar to the collective
state apearing in nuclei \cite{gdrnuclei} and metallic clusters
\cite{gdrclusters}. In paper \onlinecite{PG99}, the Bethe-Goldstone 
equations (the independent-pair approximation in Nuclear Physics
\cite{BG}) are applied to study small ($2\le N\le 6$) clusters.
The frequency for optical absorption with creation of an
electron-hole pair shows a very interesting behaviour 
related to the apparent instability of the free (not confined) 
four-exciton cluster in two dimensions.

\section{Perturbation theory}

We study a direct-band-gap semiconductor with two parabolic
bands. $N$ electrons and $N$ holes are created by, e. g., strong 
laser pumping. We shall ignore recombination processes and
electron-hole exchange. A model like this have been employed
for the analysis of collective excitations in bulk 
semiconductors \cite{CG88}.
The particles are forced to move in a two-dimensional region 
confined by a parabolic potential. This is a common approach
in the study of self-assembled quantum dots \cite{WHFJ96}.
For simplicity, we take $m_h=m_e$ and the same confining potential 
for both particles. Up to 210 electron-hole pairs will be allowed
in the dot.

The Hamiltonian of the system in oscillator units is written as

\begin{equation}
\frac{H}{\hbar\omega}=\frac{1}{2}\sum_{\alpha=1}^{2 N}
 (\vec p_{\alpha}^{~2}+ \vec r_{\alpha}^{~2})+\beta 
 \sum_{\alpha<\gamma}\frac{q_{\alpha} q_{\gamma}}
 {|\vec r_{\alpha}- \vec r_{\gamma}|},
\label{eq1}
\end{equation}

\noindent
where $\omega$ is the dot frequency, $q_{\alpha}=-1$ for 
electrons and +1 for holes. This Hamiltonian depends only 
on one constant, $\beta=\sqrt{(\frac{m e^4}{\kappa^2\hbar^2})/
(\hbar\omega)}=\sqrt{E_c/(\hbar\omega)}$, where $E_c$ is the 
Coulomb characteristic energy. $m$ is the electron effective 
mass, and $\kappa$ the dielectric constant of the material. 
In these units, the effective Bohr radius is $a_B=1/\beta$.

$\beta\to 0$ is a high-density (strong confinement) limit 
in which the Bohr radius is 
much higher than the oscillator length (equal to one in our units).
The independent-electron and -hole picture works in this limit, and
the Coulomb interaction may be computed in perturbation theory. 
Notice that at high density, we have a system of independent 
fermions, not bosons. This is one of the reasons preventing Bose
condensation of excitons in homogeneous 3D systems\cite{Keldysh}.

On the other hand, as $\beta$ is increased, the dynamics become 
more and more dictated by the Coulomb forces. First, we shall 
observe the emergence of two-body correlations and the formation
of electron-hole ``Cooper'' pairs (i. e. pairing in Fock space). 
With a further increase in $\beta$, small excitons,  
biexcitons and higher complexes shall start playing a dominant
role.

Let us first consider the $\beta\to 0$ limit, in which the Coulomb 
interaction may be computed in perturbation theory.  We will
make an additional simplifying assumption: $N$ is such that there are 
$N_{shell}$ closed shells in the $\beta=0$ limit, that is the number
of electrons takes one of the following values $N=N_{shell}
(N_{shell}+1)=$ 6, 12, 20, 30, 42, \dots, 210. The ground state of such 
systems for small $\beta$ values is spin-unpolarised, which means that 
both the electron and hole subsystems have total spin $S=0$. The 
angular momentum of this state is $L=0$.

The $\beta\to 0$ perturbative series take the form

\begin{equation}
\frac{E}{\hbar\omega}= b_0 +b_1 \beta +b_2 \beta^2 +
 {\cal O}(\beta^3),
\label{eq3}
\end{equation}

\noindent
where, the leading approximation to the energy is twice the 
energy of $N$ independent electrons (or holes)

\begin{equation}
b_0=2 N\sqrt{4 N+1}/3,
\end{equation}

\noindent
and for $b_1$ and $b_2$ we arrive to the following 
expressions

\begin{eqnarray}
b_1= - 2 \sum_{n_1\le N/2}\langle n_1,n_1|1/r|n_1,n_1\rangle 
 - 4\sum_{n_1<n_2\le N/2}\langle n_1,n_2|1/r|n_2,n_1\rangle,
\end{eqnarray}

\begin{eqnarray}
b_2= &-& 4\sum_{n_1\le N/2}\sum_{n_2>N/2}\frac{\left |
 \sum_{n\le N/2}\langle n_2,n|1/r|n,n_1\rangle\right |^2}
 {\epsilon(n_2)-\epsilon(n_1)} 
      - 6\sum_{n_1\le N/2}\sum_{n_3>N/2}\frac{\langle n_3,n_3|
 1/r|n_1,n_1\rangle^2}{\epsilon(n_3)-\epsilon(n1)}\nonumber\\
     &-& 24\sum_{n_1<n_2\le N/2}\sum_{n_3>N/2}\frac{\langle 
 n_3,n_3|1/r|n_1,n_2\rangle^2}{2\epsilon(n_3)-\epsilon(n1)-
 \epsilon(n_2)}
      - 24\sum_{n_1\le N/2}\sum_{n_4>n_3>N/2}\frac{\langle 
 n_3,n_4|1/r|n_1,n_1\rangle^2}{\epsilon(n_3)+\epsilon(n4)-
 2\epsilon(n_1)}\nonumber\\
&-& 8\sum_{n_1<n_2\le N/2}\sum_{n_4>n_3>N/2}\left\{ 5\langle 
 n_3,n_4|1/r|n_1,n_2\rangle^2\right. \nonumber\\
     &+& 5\langle n_3,n_4|1/r|n_2,n_1\rangle^2 
      - \left.4 \langle n_3,n_4|1/r|n_1,n_2\rangle
 \langle n_3,n_4|1/r|n_2,n_1\rangle \right\} \nonumber\\
     &\times& \left (\epsilon(n_3)
 +\epsilon(n_4)-\epsilon(n1)-\epsilon(n_2)\right )^{-1}.
\end{eqnarray}

\noindent
The sums run over orbitals, which have been numbered 
sequentially. The sums over spin degrees of freedom have 
been explicitly evaluated. The Coulomb matrix elements are 
defined as

\begin{eqnarray}
\langle n_1,n_2|1/r|n_3,n_4\rangle =&\int& 
 \frac{{\rm d^2}r_1 {\rm d^2}r_2}{|\vec r_1-\vec r_2|}
 \phi_{n_1}^*(\vec r_1) \phi_{n_2}^*(\vec r_2) 
 \phi_{n_3}(\vec r_1) \phi_{n_4}(\vec r_2).
\end{eqnarray}

\noindent
The explicit form of the harmonic-oscillator orbitals is

\begin{equation}
\phi_{k,l}=C_{k,|l|} r^{|l|} L_k^{|l|}(r^2)e^{-r^2/2}e^{i l \theta},
\end{equation}

\noindent
where $C_{k,|l|}=\sqrt{k!/[\pi~(k+|l|)!]}$, and $n=(k,l)$ is a composed
index. The energy corresponding to $\phi_{k,l}$ is 
$\epsilon(k,l)=1+2 k+|l|$. In terms of these energies, we have 
$b_0=4 \sum_n~\epsilon(n)$. 

Numerical values for the coefficients $b_1$ and $b_2$ are presented in
Table \ref{table1}. The sums entering the $b_2$ coefficients were 
evaluated with
a maximum of 20 shells. With respect to the number of shells included 
in the calculations, the convergence is slow, thus we used Shank 
extrapolants\cite{Shank} to accelerate convergence. Fortunately, as a 
function of $N$, $b_2$ saturates very fast and there is no need to 
perform calculations for $N>42$. 
Notice the scaling laws $b_0\approx \frac{4}{3} N^{3/2}$, 
$b_1\approx -0.96~N^{5/4}$, $b_2\approx -1.65~N$ for $N\ge 42$. $b_1$ 
depends weaker on  $N$ (as compared with electrons, for which the power 
is 7/4 instead of 5/4 \cite{GPP97}) because of the partial cancellation
between attractive and repulsive Coulomb matrix elements. 

Least-squares fits to the data in Table \ref{table1} lead to

\begin{eqnarray}
\frac{b_1}{N^{5/4}} &=& -0.960853-\frac{0.355004}{N}+
 \frac{1.24679}{N^2}-\frac{3.37305}{N^3},\\
\frac{b_2}{N} &=& -1.665+\frac{0.568}{N}-\frac{0.313}{N^2}.
\end{eqnarray}

\subsection{Approximate scaling and breakdown of the perturbative
expansion}

For large enough $N$, we can use the asymptotic expressions 
for the coefficients to 
show that $E/N^{3/2}$ is approximately a function 
of the combination $N^{-1/4}\beta$. 

\begin{equation}
\frac{E}{\hbar\omega N^{3/2}}\approx f(\beta/N^{1/4}).
\label{scaling}
\end{equation}

The physics behind the scaling (\ref{scaling}) is the following. 
As a result of cancellation 
between Coulomb attraction and repulsion, the size of the 
system is practically constant. Then, by increasing $N$, 
we increase the density and depress the effects of the 
Coulomb interaction. Approximate scaling of the energy is 
also characteristic of
confined electron systems \cite{GPP97} and charged bosons
in two dimensions \cite{GPP99}.
Notice that in a pure electron system, where the
interparticle potential is always repulsive,
an increase in $N$ leads 
to a decrease of the density and an enhancement of 
correlation effects. The energy turns out to be a function of
$N^{1/4}\beta$ at low $\beta$.

We shall stress that the scaling law (\ref{scaling}) is
expected to be observed also in the strong coupling, 
$\beta\to\infty$, limit in which the energy shall be roughly
proportional to the energy of $N$ independent excitons,

\begin{equation}
\left. \frac{E}{\hbar\omega}\right|_{\beta\to\infty}
 =a_0 \beta^2+\cdots,
\label{eq11}
\end{equation}

\noindent
where $a_0\approx -N$. The right hand side of Eq. (\ref{eq11})
may thus be written as $N^{3/2}(-\beta^2/N^{1/2})$.
The variational results of the next sections also support
the scaling behaviour (\ref{scaling}).

A naive estimation of the convergence radius for the series 
(\ref{eq3}) gives $\beta<\beta_c=b_1/b_2$. This estimation
may be obtained formally as the pole of the Pad\'e 
approximant

\begin{equation}
P_{1,1}(\beta)=b_0+\frac{b_1 \beta}{1+q_1 \beta}, 
 ~~q_1=-b_2/b_1,
\end{equation}

\noindent
which reproduces the expansion (\ref{eq3}) for small 
$\beta$ values. Notice the high-$N$ asymptotic behaviour, 
$\beta_c\sim 0.58~ N^{1/4}$. $\beta_c$ 
gives an estimate for the density at which exciton 
effects become important. Indeed, the density in our
units is $\rho\approx N/(\pi \langle r^2\rangle)\approx
3~N^{1/2}/(2\pi)$, thus $\beta_c$ may be expressed in terms
of $\rho$. Turning back to ordinary units, we get a critical
density, $\rho_c\sim 1/(0.24~\pi~a_B^2)$, i. e. approximately 
four times the close-packing 
density of excitons, $1/(\pi~a_B^2)$. This fact is 
consistent with the belief that screening is less 
effective in two dimensions. Below 
$\rho_c$, exciton effects shall dominate the quantum 
dynamics. As will be seen, $\rho_c$ is also at the onset
of pairing in the BCS estimate of the next sections.

In the following sections, we will perform variational 
estimations expected to be valid when pairing is not
so strong, that is in the regime $\beta/N^{1/4}\le 1$.

\section{Variational estimations}

Let us turn to the variational calculations. 
The simplest variational estimation one can try is first-order 
perturbation theory. 

\begin{equation}
\frac{E}{\hbar\omega}< E_{PT1}(\beta)= b_0+b_1 \beta.
\end{equation}

This estimate may be improved by 
introducing a frequency, $\Omega$, as an additional variational
parameter, i.e. by taking as trial function the product
of two Slater determinants of harmonic-oscillator states with
a frequency $\Omega$. The result is, 

\begin{equation}
\frac{E}{\hbar\omega}< {\rm min}_{\Omega} \left\{ \frac{1}{2}(\Omega+
 1/\Omega)~E_{PT1}\left(\frac{2\sqrt{\Omega}}{\Omega+1/\Omega}\beta
 \right)\right\}.
\label{eq11a}
\end{equation}

We checked that the result coming from (\ref{eq11a}) practically
coincides with the Hartree-Fock (HF) energy for this system\cite{GC98}. 
Thus, we will call (\ref{eq11a}) the HF estimate. 

The mechanism by which the energy is lowered is pairing. We may
take account of it with the help of a BCS-like wave function\cite{E92}. 
This may be a good estimation for weak pairing, when correlations are
not so strong. In the $\beta$ axis, it means $\beta/N^{1/4}<1$.
The wave function is given by 

\begin{equation}
|BCS\rangle =\prod_{j=1}^{N_{max}} (u_j+v_jh_j^+e_{j'}^+)~
 |0\rangle_h~|0\rangle_e.
\label{eq12a}
\end{equation}

\noindent
$h_j^+$ and $e_j^+$ are hole and electron (harmonic oscillator)
creation operators acting on their respective vacua
$|0\rangle_h$ and $|0\rangle_e$. $j=(k,l,s_z)$ is a composed
index, $j'=(k,-l,-s_z)$. $s_z$ is the spin projection. $v_j$ 
and $u_j$ are normalised according to $u_j^2+v_j^2=1$. 
The total angular momentum corresponding to $| BCS\rangle$ is 
zero because the angular momentum of each pair is zero. The mean 
value of the total electron (hole) spin may be forced to be 
zero by requiring $v(k,l,s_z)=v(k,l,-s_z)$. Thus, $v_j$ 
does not depend on $s_z$ and we can write $v_n$ instead of 
$v_j$.

$|BCS\rangle$ is not an eigenfunction of the particle number
operator. In a finite system, we shall project onto the state
with the correct number of particles. This will be done in two 
steps: first, an approximate projection before variation over the
parameters $v_n$ entering the BCS function (the Lipkin-Nogami 
scheme \cite{LN}), and then an exact Monte Carlo projection of the
BCS function onto the sector with $N$ pairs \cite{CG}.

\subsection{The Lipkin-Nogami estimate}

In the Lipkin-Nogami (LN) method \cite{LN}, one assumes an 
approximate polynomial dependence of $H$ on the particle
number operator $\hat N$,

\begin{equation}
H=\lambda_0+2\lambda_1 \hat N +\lambda_2 \hat N^2.
\end{equation}

By taking expectation values of $H$ over exact and BCS
functions and comparing results, we arrive to

\begin{equation}
E_{LN}=E_{BCS}-2 \lambda_1 \left(\langle\hat N \rangle_{BCS}-
 N \right)-\lambda_2 \left(\langle\hat N^2 \rangle_{BCS}-N^2 \right),
\label{eq24}
\end{equation}

\noindent
where

\begin{eqnarray}
E_{BCS}=\langle H \rangle_{BCS} &=&
 \sum_n \{ 4\epsilon_n-2\beta\langle n,n|1/r|n,n\rangle\}
 v_n^2\nonumber\\
&-&2\beta\sum_{n_1\ne n_2}~\langle n_1,n_2|1/r|n_2,n_1\rangle
 \{v_{n_1}^2 v_{n_2}^2+v_{n_1} u_{n_1} v_{n_2} u_{n_2}\}.
\label{eq14}
\end{eqnarray}

Minimization over $\lambda_1$ leads to 

\begin{eqnarray}
N &=& \langle \sum_j~e_j^+ e_j \rangle_{BCS}
 = \langle \sum_j~h_j^+ h_j \rangle_{BCS}
 \nonumber\\
 &=2 &\sum_n~v_n^2.
\label{eqn13a}
\end{eqnarray}

The equation of minimum with respect to the $v_n$ can be written
in the form of standard gap equations

\begin{eqnarray}
\Delta_n=\beta\sum_{n_1\ne n}\langle n,n_1|1/r|n_1,n\rangle
 \frac{\Delta_{n_1}}{2\sqrt{\Delta_{n_1}^2+
 (\epsilon_{n_1}^{HF}-\mu)^2}}.
\label{delta}
\end{eqnarray}

\noindent
where the HF energies are given by

\begin{eqnarray}
\epsilon_n^{HF}=&\epsilon_n& -\frac{\beta}{2}
 \langle n,n|1/r|n,n\rangle 
 - \beta\sum_{n_1\ne n}\langle n,n_1|1/r|n_1,n\rangle~
 v_{n_1}^2-\lambda_2 (N-v_n^2),
\label{eHF}
\end{eqnarray}

\noindent
and we used the common BCS parametrization

\begin{equation}
v_n^2 = \frac{1}{2} \left(1-\frac{\epsilon_n^{HF}-\mu}
 {\sqrt{\Delta_n^2+(\epsilon_n^{HF}-\mu)^2}} \right).
\label{eq29}
\end{equation}

The chemical potential, $\mu=\lambda_1+\lambda_2/2$ was 
introduced in Eq. (\ref{eq29}). For the determination of
$\lambda_2$, the system of equations 

\begin{eqnarray}
\langle H-\lambda_0-2\lambda_1\hat N
-\lambda_2\hat N^2 \rangle_{BCS}&=&0,\\
\langle (H-\lambda_0-2\lambda_1\hat N
-\lambda_2\hat N^2)\hat N \rangle_{BCS}&=&0,\\
\langle (H-\lambda_0-2\lambda_1\hat N
-\lambda_2\hat N^2)\hat N^2 \rangle_{BCS}&=&0, 
\end{eqnarray}

\noindent
is used \cite{LN}. It makes the LN method not throughly variational. 
The first equation determines the constant $\lambda_0$. The second
turns to be equivalent to the gap equation (\ref{delta}).
For $\lambda_2$, we get

\begin{equation}
\lambda_2=\frac{a_1 a_5-a_2 a_4}{a_3 a_5-a_2^2},
\label{lambda2}
\end{equation}
 
\noindent
where

\begin{equation}
a_1=\langle H\hat N^2\rangle_{BCS}-\langle H\rangle_{BCS} 
\langle \hat N^2\rangle_{BCS},
\end{equation}

\begin{equation}
a_2=\langle \hat N^3\rangle_{BCS}-N \langle \hat N^2\rangle_{BCS}
\end{equation}

\begin{equation}
a_3=\langle H \hat N^4\rangle_{BCS}-\langle \hat N^2\rangle_{BCS}^2,
\end{equation}

\begin{equation}
a_4=\langle H \hat N\rangle_{BCS}-N \langle H \rangle_{BCS},
\end{equation}

\begin{equation}
a_5=\langle \hat N^2\rangle_{BCS}-N^2.
\end{equation}

The resulting equations were solved iteratively starting from 
$\epsilon_n^{HF}=\epsilon_n$, $\Delta_n=0.2$. First, the
explicit expressions for $v_n^2$ are used and the nonlinear
equation (\ref{eqn13a}) is solved for $\mu$. After that, we
obtain $\lambda_2$ from (\ref{lambda2}), and the $\Delta_n$ 
and $\epsilon_n^{HF}$ are recalculated from (\ref{delta},\ref{eHF}). 
The process is repeated until the variation in any of the 
$\epsilon_n^{HF}$ is less than $10^{-10}$. 

Calculations were carried out for $6 \le N\le 90$ pairs 
and a maximum of 600 one-particle states for both electrons 
and holes (i. e. 300 orbitals, because there are 2 spin
states for each orbital). The absolute error in computing
Coulomb matrix elements is less than $10^{-8}$. As is
Eq. (\ref{eq11a}), we introduced an additional parameter
$\Omega$, and used the inequality

\begin{equation}
E \le {\rm min}_{\Omega} \left\{\frac{1}{2} 
 (\Omega+1/\Omega)~E_{LN}\left(\frac{2~\Omega^{1/2}}
 {\Omega+1/\Omega} \beta \right) \right\},
\label{eq19a}
\end{equation}

\noindent
where $E_{LN}$ is the result from Eq. (\ref{eq24}) at 
$\Omega=1$.
The variation of $\Omega$ can be thought of as a simplified
self-consistent Hartree-Fock-Bogoliubov procedure,
in which the mean field is forced to be a harmonic 
potential. 

We show in Fig. \ref{fig1} the energy coming from Eq. 
(\ref{eq19a}) versus $\beta$ for 
$N=6$ and 90 pairs (the curves $LN$). The HF estimates
(\ref{eq11a}) and BCS curves (the $\lambda_2=0$ limit
of LN) are also given for comparison. The lowest curves, 
labelled ``proj'', correspond to the exact projection
of the next section. We notice that the LN method
moves the BCS threshold for pairing towards zero, but
the energy itself remains very close to the HF curve. A
significant departure occurs only for $\beta>\beta_c$. 
($\beta_c\approx 0.86$ for $N=2$, and 1.7 for $N=90$). 

\subsection{The Monte Carlo projection}

The next step is an exact projection of the wave function 
onto the $N$-pair sector. The situation is similar to the
calculations carried out for nuclei\cite{projection},
where a variety of projection methods have been developed.

We project the wave function 
after the $\{v_n\}$ are determined for given 
$\beta$ and $\Omega$. With this function, the mean value of
the hamiltonian at a shifted $\beta$ is computed and multiplied
by the factor given in Eq. (\ref{eq19a}). The projected energy 
takes the following expression,

\begin{equation}
\left. E\right|_N=\sum_{j_1,\dots,j_N}
 W(j_1,\dots,j_N)~\varepsilon (j_1,\dots,j_N),
\label{eq20}
\end{equation}

\noindent
where the $j'$s have the same meaning as in Eq. 
(\ref{eq12a}), and the sum runs over possible 
combinations of $N$
states, $\{ j_1,\dots,j_N\}$, from a maximum of 
$N_{max}$ states allowed in the LN calculation. 
The ``weights'', $W$, and ``energies'', 
$\varepsilon$, are defined as

\begin{equation}
W(j_1,\dots,j_N)=\frac{v^2_{j_1}\dots v^2_{j_N}}
 {u^2_{j_1} \dots u^2_{j_N}} \left(
 \sum_{j'_1,\dots,j'_N}
 \frac{v^2_{j'_1}\dots v^2_{j'_N}}
 {u^2_{j'_1}\dots u^2_{j'_N}}\right)^{-1},
\end{equation}

\begin{eqnarray}
\varepsilon(j_1,\dots,j_N)&=& 2 
 \sum_{j\in\{j_1,\dots,j_N\}}
 \epsilon_j-\beta \sum_{j\in\{j_1,\dots,j_N\}}
 \langle j,j|1/r|j,j \rangle 
 - \beta \sum_{j,j'\in\{j_1,\dots,j_N\},~j\ne j'}
 \langle j,j'|1/r|j',j \rangle \nonumber\\
 &-& \beta \sum_{j\in\{j_1,\dots,j_N\},~
 j'\notin \{j_1,\dots,j_N\}}
 \langle j,j'|1/r|j',j \rangle \frac{u_j v_{j'}}
 {u_{j'} v_j}.
\end{eqnarray}

The expression (\ref{eq20}) for the projected energy 
allows a simple Monte Carlo evaluation, where the sets 
$\{j_1,\dots,j_N\}$ are generated with probability
$W(j_1,\dots,j_N)$ by means of a Metropolis algorithm
\cite{GFMC}.
Other equivalent forms of Eq. (\ref{eq20}), see for example
Ref.~\onlinecite{Soloviev}, are not suited for this 
evaluation. The procedure seems to be particularly 
efficient in Nuclear Physics calculations as well\cite{CG}.

The results are also drawn in Fig. \ref{fig1}. 
The improvement is significant for $\beta\sim \beta_c$,
and its relative importance diminishes 
as $N$ is increased.

\subsection{Results}

We show in Fig. \ref{fig2} our best results for the 
energies of the systems under study. The scaled energies show 
a remarkable similarity. A significant pairing (i. e. 
departure from the HF curve) is seen only for $\beta/
N^{1/4}\ge 0.55$.

Finally, we give a parametrisation of the ground-state energy 
obtained from the best of our variational estimates. 
The energy is written in the form of a Pad\'e 
approximant\cite{icps}, 

\begin{equation}
E_{gs}=b_0+b_1\beta+\frac{b_2\beta^2+p_3\beta^3+
 p_4\beta^4}{1+q_1\beta+q_2\beta^2},
\end{equation}

\noindent
where $p_3=q1 p4/q2-b1 q2$, and the coefficients $p_4$, 
$q_1$, and $q_2$ are fitted from our numerical results.
The obtained values are shown in Table \ref{table1}.

\section{Concluding remarks}

We have studied electron-hole systems in quantum
dots under strong and intermediate confinement, 
where the dense electron-hole or the BCS
excitonic phases are present.

The breakdown of perturbation theory and 
a significant pairing in the BCS wave function,
both take place at a density which is roughly
four times the close-packing density of excitons.
We interpret this result as a crossover between 
the two phases. 

As mentioned before, with an increase in $\beta$, 
particle 
correlations shall become more and more important. 
We shall observe signals of the ``excitonic'', 
``biexcitonic'', etc. insulating  phases. The true 
energies and wave functions of these phases shall be 
obtained with more powerful methods as, for example,
Green-function Monte Carlo method\cite{GFMC}.
Even a variational Monte Carlo estimation, as 
that one carried out for the homogeneous case in
Ref.~\onlinecite{VMC}, may be biased by the 
chosen trial functions. The density matrix 
renormalisation group method of Ref.~
\onlinecite{DS99} could also be useful. 
The very interesting 
question about whether the system remains bound
after the external potential is switched off,
still remains to be answered. We have some indications
that the two-dimensional triexciton is bound and the
four-exciton system is unbound \cite{PG99}. But 
the situation may be analogous to nuclei, where
there is a small instability island around 
atomic number 5. Some of these problems are 
currently under investigation.

\acknowledgements 
The authors acknowledge support from the Colombian
Institute for Science and Technology (COLCIENCIAS).
Part of this work was done during a visit of
A. G. and B. R. to the Abdus Salam ICTP under the 
Associateship Scheme and the Visiting Young Student
Programme.

\begin{figure}
\caption{a) and b): Ground-state energies of the 6-exciton and 90-exciton 
systems respectively.}
\label{fig1}
\end{figure}

\begin{figure}
\caption{Scaling of the ground-state energies}
\label{fig2}
\end{figure}
\newpage

\begin{table}
\caption{The coefficients $b_1$, $b_2$, $p_3$, $p_4$, 
 $q_1$, and $q_2$.}
\label{table1}
\begin{tabular}{|l|l|l|l|l|l|l|}
$N$ & $b_1/N^{5/4}$ & $b_2/N$ & $p_3$ & $p_4$
& $q_1$ & $q_2$ \\
\hline\hline
6   & -1.001    & -1.58 & 21.9817  & -461906  & 15.3098 & 9.92132\\
12  & -0.983778 & -1.62 & 43.4125  & -47.3284 & 6.46992 & 4.84953\\
20  & -0.9758   & -1.64 & 53.9705  & -47.5022 & 3.60744 & 2.85761\\
30  & -0.971391 & -1.65 & 84.1615  & -64.3801 & 3.68693 & 2.58195\\
42  & -0.968681 & -1.65 & 127.946  & -115.014 & 6.85897 & 3.44577\\
56  & -0.966881 &       & 172.825  & -157.173 & 8.02801 & 3.55984\\
72  & -0.965621 &       & 213.377  & -189.426 & 8.25862 & 3.35560\\
90  & -0.9647   &       & 259.185  & -224.076 & 8.27331 & 3.16176\\
110 & -0.964009 &  & & & &\\
132 & -0.963475 &  & & & &\\
156 & -0.963055 &  & & & &\\
182 & -0.962712 &  & & & &\\
210 & -0.962433 &  & & & &\\
\end{tabular}
\label{tab1}
\end{table}

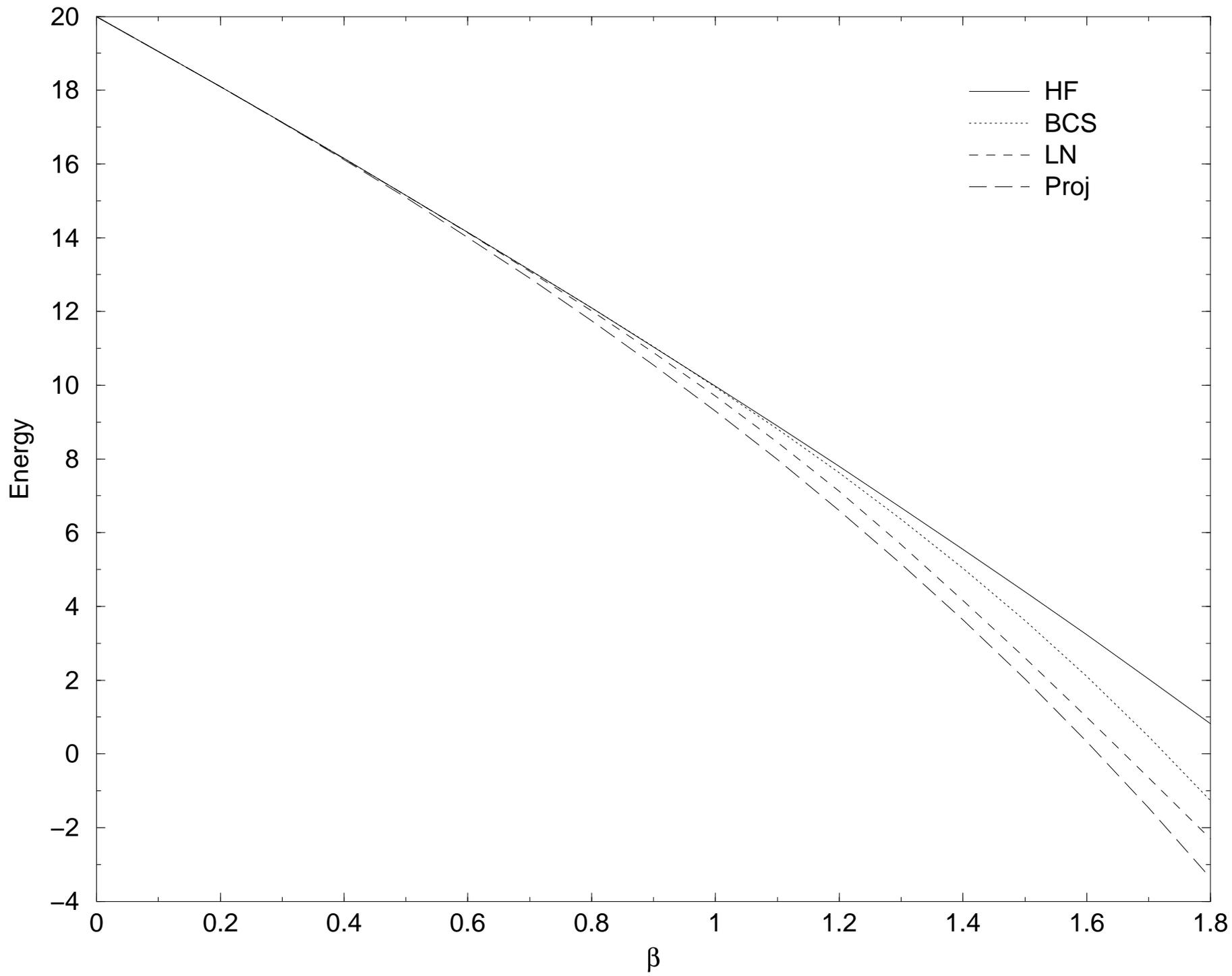
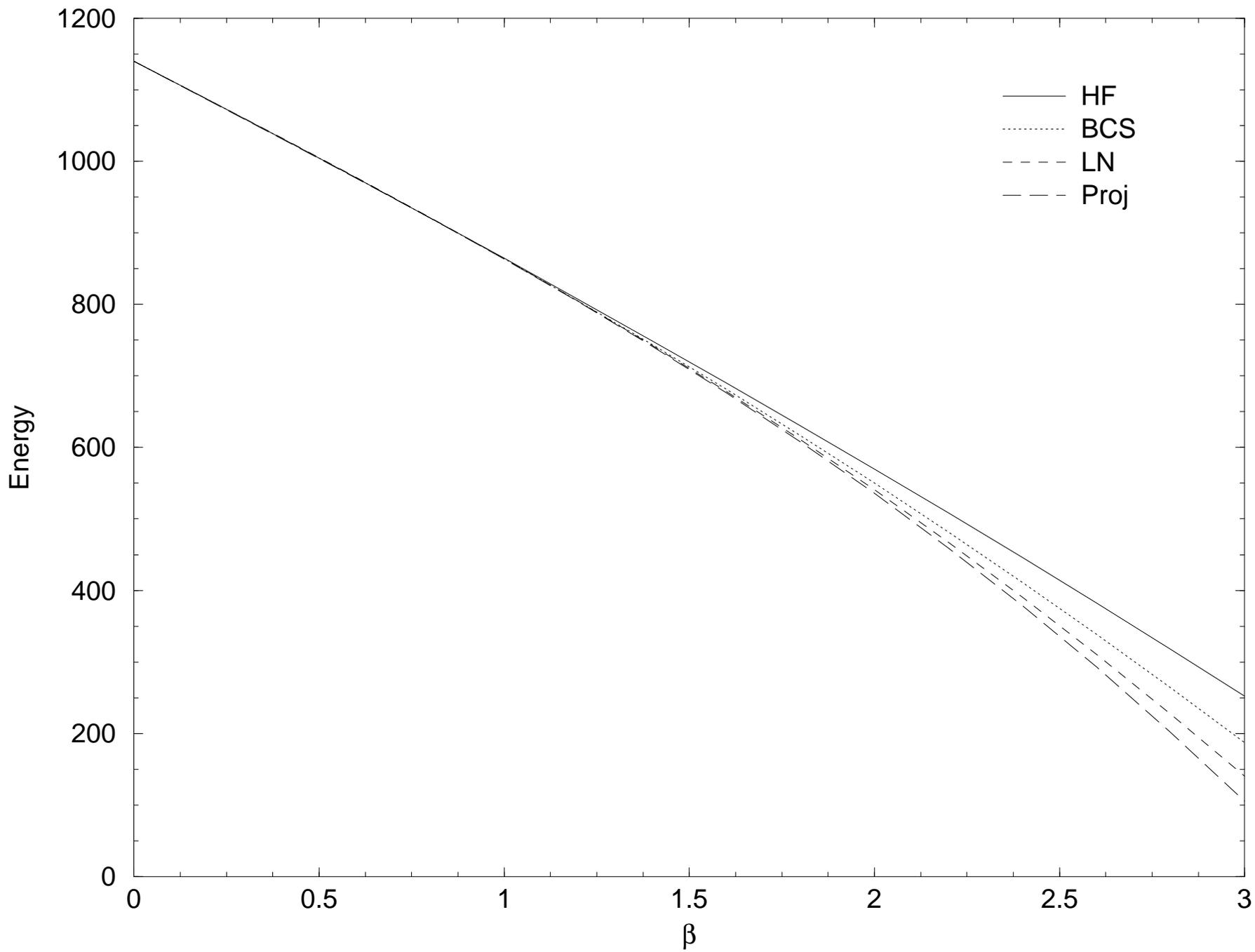
\end{document}